\documentclass{IEEEtran}

\usepackage{amsmath}
\RequirePackage{crop,graphicx,array,amsthm,stfloats,upref,amsfonts,amssymb,cuted}
\usepackage{import} 
\usepackage{graphicx, subcaption} 
\usepackage{multirow, multicol} 
\usepackage{color,soul} 

\usepackage[table]{xcolor} 
\definecolor{g}{RGB}{231, 231, 231}
\usepackage{eso-pic}
\usepackage{hyperref}
\usepackage{svg}

\usepackage{footnote}

\begin{document}

\title{\LARGE \bf
Music Boundary Detection using Convolutional Neural Networks: \break A comparative analysis of combined input features}

\author{Carlos Hernandez-Olivan, Jose R. Beltran, David Diaz-Guerra
\thanks{C. Hernández-Olivan, J.R. Beltran, and D. Diaz-Guerra are with the Department
of Electronic Engineering and Communications, University of Zaragoza,
50018 Zaragoza, Spain}
}

\maketitle

\begin{abstract}
The analysis of the structure of musical pieces is a task that remains a challenge for Artificial Intelligence, especially in the field of Deep Learning. It requires prior identification of the structural boundaries of the music pieces, whose structural boundary analysis has recently been studied with unsupervised methods and supervised neural networks trained with human annotations. The supervised neural networks that have been used in previous studies are Convolutional Neural Networks (CNN) that use  Mel-Scaled Log-magnitude Spectograms features (MLS), Self-Similarity Matrices (SSM) or Self-Similarity Lag Matrices (SSLM) as inputs. In previously published studies, pre-processing is done in different ways using different distance metrics, and different audio features are used for computing the inputs, so a generalised pre-processing method for calculating model inputs is missing. The objective of this work is to establish a general method to pre-process these inputs by comparing the results obtained by taking the inputs calculated from different pooling strategies, distance metrics and audio characteristics, also taking into account the computing time to obtain them. We also establish the most effective combination of inputs to be delivered to the CNN to provide the most efficient way to extract the boundaries of the structure of the music pieces. With an adequate combination of input matrices and pooling strategies, we obtain an accuracy $F_1$ of 0.411 that outperforms a current work done under the same conditions (same public available dataset for training and testing).
\end{abstract}

\IEEEpeerreviewmaketitle

\section{Introduction}
Music Information Retrieval (MIR\footnote{\url{https://musicinformationretrieval.com/index.html}}) is the interdisciplinary science for retrieving information from music. MIR is a field of research that faces different tasks in automatic music analysis, such as pitch tracking, chord estimation, score alignment or music structure detection. One of the most active communities and references in MIR is the Music Information Retrieval Evaluation eXchange (MIREX\footnote{\url{https://www.music-ir.org/mirex/wiki/MIREX_HOME}}). This is the community that every year holds the  International Society for Music Information Retrieval Conference (ISMIR). Algorithms are submitted to be tested in MIREX's datasets within the different MIR tasks. Most of the previous results analyzed and compared in this work have been presented in different MIREX campaigns.

The automatic structural analysis or Music Structure Analysis (MSA) of music is a very complex challenge that has been studied in recent years \cite{review}, but it has not yet been solved with an adequate accuracy that surpasses the analysis performed by musicians or specialists. This kind of analysis is only a part of the musical analysis, which involves musical aspects like harmony, timbre and tempo, and segmentation principles like repetition, homogeneity and novelty \cite{muller2015fundamentals}.
This automatic music analysis can be faced starting from music representations such as the score of the piece, the MIDI file of the piece, or the raw audio file.

In music, \textit{form} refers to the structure of a musical piece, which consists of dividing the musical pieces into small units, starting with the motifs, then the phrases, and finally the sections that express a musical idea. \textit{Boundary detection} is the first step that has to be done in musical form analysis and must be done before the naming of the different segments depending on the similarity between them. This last step is named \textit{Labelling} or \textit{Clustering}. This task, translated to the most common genre in MIREX datasets, the pop genre, would be the detection and extraction of the chorus, verse, or introduction of the corresponding song. Detecting the boundaries of music pieces consists on identifying the transitions where these parts begin and end, a task that professional musicians do almost automatically by listening a piece of music. This detection of the boundaries in a musical piece is based on the \textit{Audio Onset Detection} task, which is the first step for several higher-level music analysis tasks such as beat detection, tempo estimation, and transcription.

This problem can be accomplished with different techniques that have in common the need of pre-processing the audio files in order to extract the desired audio features and then apply unsupervised or supervised methods. There are several studies where this pre-processing step is made in different ways, so there is not yet a generalized input pre-processing method. The currently \textit{end-to-end} best-performing methods use CNNs trained with human annotations. The inputs to the CNN are Mel-Scaled Log-magnitude Spectograms (MLSs) \cite{ullrich2014boundary}, Self-similarity Lag-Matrices (SSLMs) in combination with the MLSs \cite{grill2015music}, and also combining these matrices with chromas \cite{grill2015music2}.

One of the limitations of these methods is that the analysis and results obtained depend largely on the database annotator since there can be inconsistencies between different annotators when analyzing the same piece. Therefore, these methods are limited to the quality of the labels given by the annotators and they cannot outperform them.
 
This paper deals with the issue of structure detection in music pieces. In particular, we study the comparison of different methods of boundary detection between the musical sections by means of Convolutional Neural Networks. The paper is structured as follows: Section \ref{sec:related_work} presents an overview of the related work and previous studies in which this work is based on. The Self-Similarity Matrices and the used datasets are also presented. In Section \ref{sec:method}, the pre-processing method of the matrices that will be used as inputs of the neural network (NN) is explained. Section \ref{sec:dataset} introduces the database used for training, validating and testing, and the labelling process. Section \ref{sec:model} shows the NN structure and the thresholding and peak-picking strategies and section \ref{sec:results} describes the metrics used to test the model and exposes the results of the experiments and their comparison with previous studies. Finally, section \ref{sec:discussion} presents the discussion and section \ref{sec:conclusions} discusses proposals for future lines of work. All code used in this paper, including the pre-trained models of every case of study in this work, is made publicly available \footnote{\url{https://github.com/carlosholivan/MusicBoundariesCNN}} and further results are shown in the website \footnote{\url{https://carlosholivan.github.io/publications/2021-boundaries/2021-boundaries.html}}.

\section{Related Work} \label{sec:related_work}
Several studies have been done in the field of structure recognition in music since Foote introduced the self-similarity matrix (SSM) in 1999 \cite{foote1999visualizing} and later, in 2003, he derived from it the self-similarity lag matrix (SSLM) \cite{goto2003chorus}. 
Before the introduction of the SSMs and SSLMs, the studies were based on processing spectrograms \cite{zhang1999heuristic}, but in recent years it has been demonstrated that SSMs and SSLMs calculated from audio features in combination with spectrograms provide better results.
We describe some previous works of both unsupervised and supervised methods which belongs to the MIREX's task: Music Structure Segmentation.

\subsection{Unsupervised Methods} 
\label{sec:un}
The main idea of most of the unsupervised methods is to extract the musical structure of the music pieces but not necessarily the boundaries between the structure sections.

According to Paulus et al. \cite{paulus2010state}, these methods can be summarized in three approaches based on: novelty, homogeneity and repetition. These approaches are computed with unsupervised Machine Learning algorithms such as genetic algorithms (\textit{fitness functions}), Hidden Markov Models (HMM), \textit{K-means}, Linear Discriminant Analysis (NDA), Decision Stump or Checkerboard-like kernels.

The \textbf{Novelty-based} approach consists on the detection of the transitions between contrasting parts \cite{review}. This approach is well-performed using checkerboard-like kernel methods which were introduced by Foote in 2000 \cite{foote2000automatic}.
These methods have evolved during the years and it has been found that multiple-temporal-scale kernels, as those of Kaiser and Peeters in 2013 \cite{kaiser2013multiple}, outperformed the results of previous works by proposing a fusion of the novelty and repetition approaches.

The \textbf{Homogeneity-based} approach is based on the identification of sections that are consistent with respect to their musical properties \cite{review}. These methods use Hidden Markov Models, like Logan and Chu \cite{logan2000music}, Aucouturier and Sandler \cite{aucouturier2001segmentation} and Levy and Schandler \cite{levy2008structural} or combinations of SSMs like Traile and McFee \cite{tralie2019enhanced}, and McFee and Bello \cite{mcfee2017structured}.

The \textbf{Repetition-based} approach refers to finding recurring patterns. These methods apply a clustering algorithm to the SSMs or SSLMs. They are more applicable for labeling the structural parts of music pieces rather than precise segmentation which is required by boundary detection.
Lu et al. in 2004 \cite{lu2004repeating}, Paulus and Klapuri in 2006 \cite{paulus2006music}, Turnbull et al. \cite{turnbull2007supervised}, McFee and Ellis \cite{mcfee2013dp1}, and McCallum \cite{mccallum2019unsupervised} are examples of this method. 

To conclude, we can affirm that unsupervised algorithms are very efficient performing the labelling (clustering) part, but not the boundaries detection task, which is better performed by supervised neural networks which came up in 2014 and are described in section \ref{sec:endtoend}.

\subsection{Supervised Neural Networks} \label{sec:endtoend}
\begin{figure*}[ht!]
	\centering
	\centering
	    \includegraphics[scale=0.7]{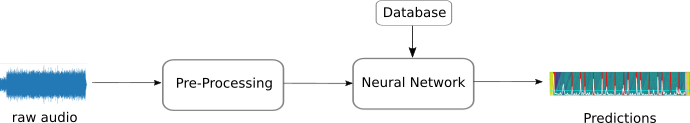}
	    \caption{General scheme of supervised neural networks.}
	    \label{general}
\end{figure*}

Supervised neural networks learn from input representations given the ground truth, which are the label annotations of the targets (Fig. \ref{general}).

Previous studies of boundary detection used Mel-Scaled Log-magnitude Spectograms (MLS) as the inputs of CNNs \cite{ullrich2014boundary}. This method was based on \textit{Audio Onset Detection} task \cite{schluter2014improved}, which consists on finding the starting points of every musically relevant event in an audio signal, specifically the beginning of a music note. This task can be interpreted as a computer vision problem, like edge detection, but applied to spectrograms instead of images with different textures.

Later on, in 2015, Grill and Schl\"uter improved their previous work by adding SSLMs, which yielded to better results \cite{grill2015music}, and the addition of SSLMs with different lag factors to the input of the CNN \cite{grill2015music2}, outperforming this method and reaching the best result to date.

In Tables \ref{all_table1} and \ref{all_table2} we show a recap of the results of almost all of the previous works that have been done in boundary detection using both unsupervised and supervised neural networks. Results and algorithms nomenclature in Table \ref{all_table1} have been extracted from MIREX's campaigns of different years. It must be said that the results obtained with unsupervised methods on Table \ref{all_table1} are not as high as the results obtained with supervised neural networks because, as it has been mentioned in section \ref{sec:un}, the main goal of the unsupervised methods is not the boundary detection (segmentation) itself but the full structure identification (labelling).

\subsection{Self-Similarity Matrices (SSMs)} \label{ssm}
The Self-Similarity Matrix \cite{muller2015fundamentals} is a tool not only used in music structure analysis but also in time series analysis tasks. In these matrices, the different parts of the structure of a music piece can be identified as homogeneous regions. This representation of the structural elements of music analysis leads this matrix and its combination with spectrograms to be the input of almost every model described in sections \ref{sec:un} and \ref{sec:endtoend}. For this work, this matrix is important because music is in itself \textit{self-similar}, in other words, it is formed by similar time series. 

Self-Similarity Matrices have been used under the name of Recurrence Plot for the analysis of dynamic systems \cite{jp1987recurrence}, but their introduction to the music domain was done by Foote \cite{foote1999visualizing} in 1999 and since then, there have appeared different techniques for computing these matrices.
The SSM relies on the concept of self-similarity, which is measured by a similarity function that is applied to the audio features representation. The similarity between two feature vectors $y_n$ and $y_m$ is a function that can be expressed as Eq. \ref{similarity} shows. The result is a $N$-square matrix SSM $\in \mathbb{R}^{N\mathrm{x}N}$ being $N$ the time dimension:

\begin{equation} \label{similarity}
\mathrm{SSM}(n,m) = \delta (y_n, y_m)
\end{equation}
where $n,m \in$ [\ 1,...,$N$]\ .\\

\begin{table*}[h!]
\renewcommand{\arraystretch}{1.2}
\caption{Results of boundary detection of previous studies for "Full Structure" and "Segmentation" tasks. Only the best-performing algorithm in terms of F-measure of each year for a $\pm$0.5s time-window tolerance is shown. The F-measure is shown for different databases (see Sec.\ref{data}).} 
\label{all_table1}
\centering
\begin{tabular}{p{0.5cm} p{2.5cm} p{1cm} >{\centering\arraybackslash}
p{2cm} >{\centering\arraybackslash}p{1.7cm} | c c c c}
\hline
\hline
\multicolumn{9}{c}{\textbf{Unsupervised Methods}}\\
\hline
\multirow{2}{*}{\textbf{Year}\footnote} &
\multirow{2}{*}{\textbf{Autors [Ref.]}} &
\multirow{2}{*}{\textbf{Algorithm}} &
\multirow{2}{*}{\textbf{Input}} &
\multirow{2}{*}{\textbf{Method}} &
\multicolumn{4}{c}{\textbf{F-measure (\textbf{$\mathrm{F_{1}}$}) for Testing Databases}}\\
 &  &  & & & \textbf{MIREX09} & \textbf{RCW-A} & \textbf{RCW-B} & \textbf{SALAMI}\\

\hline
2009 & Paulus \& Klapuri \cite{320092} & PK & MFCCs, chromas & \textit{Fitness function} & 0.27 & - & - & -\\

2010 & Mauch et al. \cite{mauch2009using} & MND1 & MFCCs, Discrete Cepstrum & \textit{HMM} & 0.325 & 0.359  & - & -\\ 

2011 & Sargent et al. \cite{sargent:hal-00618141} & SBVRS1 & Chords estimation & \textit{Viterbi} & 0.231 & 0.324 & - & -\\ 

2012 & Kaiser et al. \cite{kaiser2012mirex} & KSP2 & SSM & \textit{Novelty measure} & 0.280 & 0.366 & 0.289 & 0.286\\

2013 & McFee \& Ellis \cite{mcfee2013dp1} & MP2 & MLS & \textit{Fisher's Linear Discriminant} & 0.281 & 0.355 & 0.278 & 0.317\\

2014 & Nieto \& Bello \cite{nieto2014mirex} & NB1 & MFCCs + chromas & \textit{Checkerboard-like kernel} & 0.289 & 0.352 & 0.269 & 0.299\\

2015 & Cannam et al. \cite{cannam2015mirex} & CC1 & Timbre-type histograms & \textit{HMM} & 0.197 & 0.224 & 0.203 & 0.213\\

2016 & Nieto \cite{nietomirex} & ON2 & Constant-Q Transform Spectrogram & \textit{Linear Discriminant Analysis} & 0.259 & 0.381 & 0.255 & 0.299\\

2017 & Cannam et al. \cite{cannam2015mirex} & CC1 & Timbre-type histograms & \textit{HMM} & 0.201 & 0.228 & 0.192 & 0.212\\

\hline
\hline
\multicolumn{9}{c}{\textbf{Supervised Neural Networks}} \\

\hline
2014 & Schlüter et al. \cite{schluter2014structural} & SUG1 & MLS & CNN & 0.434 & 0.546 & 0.438 & 0.529\\
2015 & Grill \& Schlüter \cite{grill2015structural} & GS1 & MLS + SSLMs & CNN & 0.523 & 0.697 & 0.506 & 0.541\\

\hline
\hline
\end{tabular}
\end{table*}
\footnotetext{\url{https://www.music-ir.org/mirex/wiki/<<year>>:MIREX<<year>>_Results} - headland ''Music Structure Segmentation Results''.}

The similarity function is obtained by the calculation of a distance between the two feature vectors $y$ mentioned before. In the literature, this distance is usually calculated as the Euclidean distance $\delta_{eucl}$ or the cosine distance $\delta_{cos}$:

\begin{equation} \label{distance_euc}
    \mathrm{\delta_{eucl}}(y_n, y_m) = \Vert y_n - y_m \Vert
\end{equation}
\begin{equation} \label{distance_cos}
\mathrm{\delta_{cos}}(y_n, y_m) = \mathrm{1 -} \frac{u.v}{\Vert y_n \Vert . \Vert y_m \Vert}
\end{equation}
where \textit{u} and \textit{v} are time series vectors.

Self-Similarity Matrices can be computed from different audio features representations, such as MFCCs or chromas, and they can also be obtained by combining different frame-level audio features \cite{tralie2019enhanced}.
Once the similarity function has been computed for each pair of audio feature vectors and the SSM has been calculated, we can filter the SSM by applying thresholding techniques, smoothing or invariance transposition. 
The SSM can also be obtained with other techniques such as clustering methods as Serra et al. proposed \cite{serra2014unsupervised}, where the SSM is obtained by applying the \textit{k-nn} algorithm.

After Foote in 1999 defined the SSM, in 2003, Goto \cite{goto2003chorus} defined a variant of the SSM which is known as the Self-Similarity Lag Matrix (SSLM). The SSLM is a matrix that represents the similarities between low-level features of one point in time and points in the past, up to
a certain \textit{lag time}. This representation makes possible to plot the relations between past events and their repetitions in the future. Some approaches calculate this SSLM after computing the SSM or the recurrence plot as we show in Eq. \ref{lag}:

\begin{equation} \label{lag}
   \mathrm{SSLM}(i,j) = \mathrm{SSM}_{k+1,j} 
\end{equation}

with $i = 1, ... , N$, $j = 1, ... , L$ and $k = i + j - 2 modulus (N)  $\\

\begin{table*}[h]
\caption{Results of previous works in boundary detection task for $\pm$0.5s time-window tolerance. It is only showed the best F-measure result of each reference for each database.} \label{all_table2}
\centering
\begin{tabular}{p{0.5cm} p{2cm} p{1.8cm} c c | c c c c}
\hline
\hline
\multicolumn{9}{c}{\textbf{Unsupervised Methods}}\\
\hline
\multirow{2}{*}{\textbf{Year}} &
\multirow{2}{*}{\textbf{Autors [Ref.]}} &
\multirow{2}{*}{\textbf{Input}} &
\multirow{2}{*}{\textbf{Method}} &
\multirow{2}{*}{\textbf{Train Set}} &
\multicolumn{4}{c}{\textbf{F-measure (\textbf{$\mathrm{F_{1}}$}) for different Databases}}\\
& &  &  & &\textbf{MIREX09} & \textbf{RCW-A} & \textbf{RCW-B} & \textbf{SALAMI}\\
\hline
2007 & Turnbull et al. \cite{turnbull2007supervised} & MFCCs, chromas, spectrogram & \textit{Boosted Decision Stump} & - & - & - & 0.378 & -\\
2011 & Sargent et al. \cite{sargent2011regularity} & MFCCs,chromas & \textit{Viterbi} & - & - & - & 0.356 & -\\

\hline
\hline
\multicolumn{9}{c}{\textbf{Supervised Neural Networks}} \\
\hline
2014 & Ullrich et. al \cite{schluter2014improved} & MLS & CNN & Private  & - & - & - & 0.465\\
2015 & Grill \& Schlüter \cite{grill2015music} & MLS + SSLMs & CNN & Private & - & - & - & 0.523\\
2015 & Grill \& Schlüter \cite{grill2015music2} & MLS + PCPs + SSLMs & CNN & Private & - & - & - & 0.508\\
2017 & Hadria \& Peeters \cite{cohen2017music} & MLS + SSLMs & CNN & SALAMI & - & - & - & 0.291\\
\hline
\hline
\end{tabular}
\end{table*}

The dimensions of this matrix are not $N\times N$ as the SSM, but they are $N\mathrm{x}L$, being $L$ the \textit{lag time factor}. That means that the SSLM is a non-square matrix: SSLM $\in \mathbb{R}^{N \times L}$.

The choice of the type of audio features representation for computing the SSMs or SSLMs, and the choice of using SSMs or SSLMs is one of the most important steps when solving a MIR task and has to be studied depending on the issue we we want to face.

\subsection{Datasets} \label{data}
Previous works had been tested in the annual Music Information Retrieval Evaluation eXchange (MIREX \cite{downie2010music}), which is a framework for evaluating music information retrieval algorithms.

The first dataset of the MIREX campaign for  the structure segmentation task was the MIREX09 dataset, consisting on a collection of The Beatles' songs plus another smaller dataset\footnote{\url{http://ifs.tuwien.ac.at/mir/audiosegmentation.html}}. Beatles dataset have 2 annotation versions, one is Paulus Beatles or Beatles-TUT\footnote{\url{http://www.cs.tut.fi/sgn/arg/paulus/beatles_sections_TUT.zip}} dataset and the second one is the Isophonic Beatles or Beatles-ISO\footnote{\url{http://isophonics.net/content/reference-annotations}} dataset. The second MIREX dataset was MIREX10, formed by the RWC \cite{goto2004development} dataset. This dataset has 2 annotation versions; RWC-A\footnote{\url{http://musicdata.gforge.inria.fr}} of QUAERO project which is the one which corresponds to MIREX10 and RWC-B\footnote{\url{http://staff.aist.go.jp/m.goto/RWC-MDB/AIST-Annotation}} \cite{goto2006aist}, which is the original annotated version following the annotation guidelines established by Bimbot el al. \cite{bimbot2012methodology}.

A few years later, the MIREX12 dataset provided a greater variety of songs than the MIREX10 \cite{ehmann2011music}. MIREX12 is a dataset formed by the ''Structural Analysis of Large Amounts of Music Information'' (SALAMI\footnote{\url{https://ddmal.music.mcgill.ca/research/SALAMI/}}) dataset which has evolved in its more recent version, the SALAMI 2.0 database. The analysis of MIREX structure segmentation task was published in 2012 \cite{smith2013meta}.
Our work uses the publicly available SALAMI 2.0 dataset.

\section{Audio Processing} \label{sec:method}
This work is based on the previous works of Schuler, Grill et al. \cite{ullrich2014boundary}, \cite{grill2015music} who propose a pre-proscessing method to obtain the SSLMs from MFCCs features. We will extend these works by calculating the SSLMs from chroma features and applying also the Euclidean distance that has not been considered in preliminary studies, to compute the SSLMs in order to give a comparison and find the best-performing input to the NN model.

\subsection{Mel Spectrogram} \label{mel}
The first step of the pre-processing part is to extract the audio features. To do that, we first compute the the Short-Time-Fourier-Transform (STFT) with a Hanning window of 46ms (2048 samples at 44.1kHz sample rate) and an overlap of 50\% as Grill et al. proposed \cite{grill2015music}. Then, we obtain a mel-scaled filterbank of 80 triangular filters from 80Hz to 16kHz and we scale logarithmically the amplitude magnitudes to obtain the mel-spectrogram (MLS). We used the \textit{librosa} library \cite{mcfee2015librosa} to compute the mel-spectrogram. After obtaining the MLS, we apply a max-pooling of \textit{p} = 6 in the temporal dimension to give the Neural Network a manageable size input. The size of the MLS matrix is $P \times N$ with $P$ being the number of frequency bins (that are equal to the number of triangular filters) and $N$ the number of time frames. We define $\mathbf{x}_i$ with $i=1 \ldots N$ as the $i$-th frame of the MLS.

\subsection{Self-Similarity Lag Matrix from MFCCs} 
\label{sslm_mfccs}

The method that we used to generate the SSLMs\footnote{\url{https://github.com/carlosholivan/SelfSimilarityMatrices}} is the same method that Grill and Schluter used in \cite{grill2015music} and \cite{grill2015music2}, which in turn derives from Serra et al. \cite{serra2012unsupervised}.

The first step after computing each frame mel-spectrogram $\mathbf{x}_i$ is to pad a vector $\Phi$ with noise of -70dB with a duration of $L$ frames at the beginning of the mel-spectrogram.

\begin{equation} \label{pad}
\check{\mathbf{x}}_i = \Phi \mathbin\Vert \mathbf{x}_i
\end{equation}
where $\Phi$ is a matrix of size $L \times P$ whose elements are equal to -70dB.

Then, a max-pool of a factor of $p_1$ is done in the time dimension as shown in Eq. \ref{max_pool}.
\begin{equation} \label{max_pool}
\mathbf{x}_i^\prime = \max_{j=1 \ldots p_1}(\check{\mathbf{x}}_{(i-1)p_1+j})
\end{equation}

After that, we apply a Discrete Cosine Transform of Type II to each frame omitting the first element.
\begin{equation} \label{DCT}
\tilde{\mathbf{X}}_i = \mathrm{DCT}^{\mathrm{(II)}} (\mathbf{x}_i^\prime)_{[2 \ldots P]}
\end{equation}
where $P$ are the number of mel-bands.

Now we stack the time frames by a factor $m$ so we obtain the time series in Eq. \ref{3}. The resulting $\hat{\mathbf{X}}_i$ vector has dimensions $[(P-1)\cdot m]\times [(N+L)/p_1$] where $N$ is the number of time frames before the max-pooling and $L$ the lag factor in frames.
\begin{equation} \label{3}
\hat{\mathbf{X}}_i = [\tilde{\mathbf{X}}_i^\mathrm{T} \mathbin\Vert \tilde{\mathbf{X}}_{i+m}^\mathrm{T}]^\mathrm{T}
\end{equation}

The final SSLM matrix is obtained by calculating a distance between the vectors $\hat{\mathbf{X}}_i$. In our work, we use two different distance metrics: the Euclidean distance and the cosine distance. This will allow us to make a comparison between them and conclude which SSLM performs better. 

Therefore, the distance between two vectors $\hat{\mathbf{X}}_i$ and $\hat{\mathbf{X}}_{i-l}$ using the distance metric $\delta$ is
\begin{equation} \label{4}
D_{i,l} = \delta(\hat{\mathbf{X}}_i,\hat{\mathbf{X}}_{i-l}), \quad l = 1 \ldots \left\lfloor \frac{L}{p_1}\right \rfloor
\end{equation}
where $\delta$ is the distance metric as defined in Eqs. \ref{distance_euc} and \ref{distance_cos}. 

Then, we compute an equalization factor $\varepsilon_{i,l}$ with a quantile $\kappa$ of the distances $\delta(\hat{\mathbf{X}}_i,\hat{\mathbf{X}}_{i-j})$ for $j = \mathrm{1} \ldots \left\lfloor \frac{L}{p}\right \rfloor$

\begin{equation} \label{epsilon}
\varepsilon_{i,l} = Q_\kappa \left( D_{i,l}, \cdots ,D_{i,\left\lfloor\frac{L}{p}\right \rfloor} \mathbin\Vert D_{i-l,1}, \cdots ,D_{i-l,\left\lfloor\frac{L}{p}\right \rfloor} \right)
\end{equation}

We now remove the first $L/p$ lag bins in the time dimension of the distances matrix $D$ and in the equalization factor matrix $\varepsilon$, and we apply Eq. \ref{max_pool} with max-pooling factor $p_2$. Finally we obtain the SSLM applying Eq. \ref{R}.

\begin{equation} \label{R}
R_{i,l} = \sigma\left(\mathrm{1}-\frac{D_{i,l}}{\varepsilon_{i,l}}\right)
\end{equation}
where $\sigma(x) = \frac{1}{1+e^{-x}}$.\\

Once the SSLM has been obtained, we need to pad some noise to the begin and end of the SSLM because the labels which are used to train our model will be given to the NN as Gaussians (see section \ref{sec:dataset}), so the first and last labels need information in their left and right sides respectively.
We add the noise to the begin and end of the SSLM and MLS by padding them with $\gamma$ = 50 time frames of pink noise at the beginning and end of the MLS matrix. Then we then normalized each frequency band to zero mean and unit variance for MLS and each lag band for the SSLMs. Note also that if there are some time frames that have exactly the same values, the cosine distance would give a NAN (not-a-number) value. We avoid this by converting all this NAN values into zero as the last step of the SSLM computation.  

\subsection{Self-Similarity Lag Matrix from Chromas} \label{sslm_chromas}
The process of computing the SSLM from chroma features is similar to the method explained in section \ref{sslm_mfccs}. The difference here is that instead of starting with padding the mel-spectrogram in Eq. \ref{pad}, we pad the STFT. After applying the max-pooling in Eq. \ref{max_pool}, we compute the chroma filters instead of computing the DCT in Eq. \ref{DCT}. The rest of the process is the same as described in section \ref{sslm_mfccs}.

\begin{figure}[h!]
	\centering
	\includegraphics[width=\columnwidth]{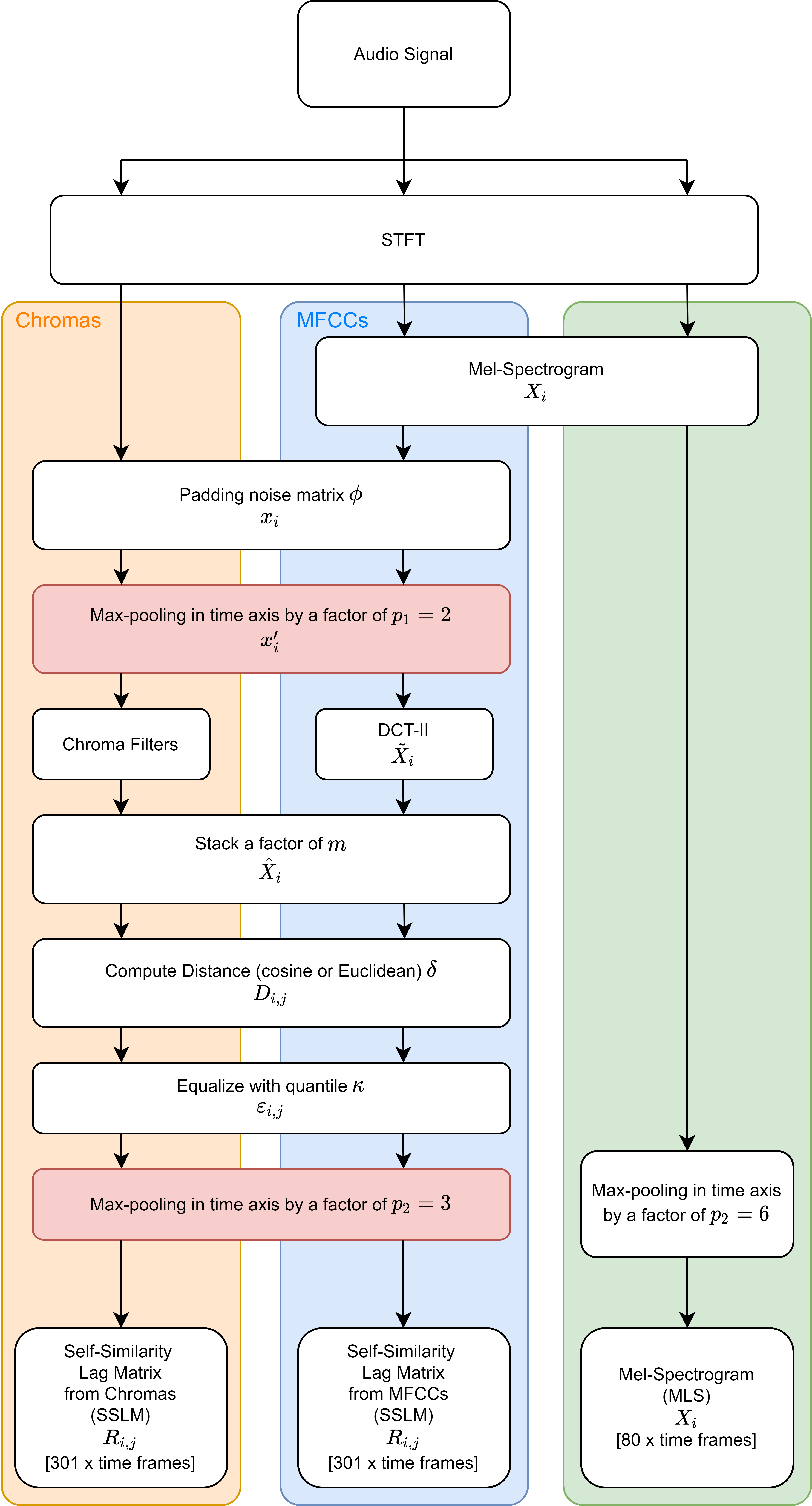}
	\caption{General block diagram of the pre-processing block in Fig. \ref{general}. Each background color contains the steps that are necessary to compute each of the inputs: MLS (green), SSLM from Chromas (orange) and SSLM from MFCCs (blue). The red background in the max-pooling blocks refers to the 2 variants done in this work: \texttt{2pool3} is the one showed in the scheme, while \texttt{6pool} is computed by applying the max-pooling of factor 6 in the first red block and removing the second red block of the scheme.}
	\label{scheme}
\end{figure}

All the values of the parameters used to obtaining the Self-Similarity Matrices are summarized in Table \ref{params}. In addition to the Euclidean and cosine metrics, and MFCCs and chromas audio features, we will compare two pooling strategies. The first one is to make a max-pooling of factor $p_1$ = 6 to the STFT (from MLS calculation), and to the Chromas or MFCCs for the SSLMs computation, as it is described in Eq. \ref{max_pool}. The other pooling strategy is the one showed in Fig. \ref{scheme}, where we first do a pooling of $p_1$ = 2 and then a pooling of $p_2$ = 3 once the SSLMs are obtained. We denote these pooling variants as \texttt{6pool} and \texttt{2pool3} respectively. The total time for processing all the SSLMs (MFCCs and cosine distance) was a factor or 4 faster for \texttt{6pool} than \texttt{2pool3} because by applying a higher padding factor in Eq. \ref{max_pool} the size of the matrices $D$ and $\varepsilon$ is much lower so the calculation of these matrices take more time but it also implies a resolution loss that can affect the accuracy of the model as \cite{grill2015music} remarks. 

The general schema of the pre-processing block is depicted in Fig. \ref{scheme}.

\begin{table}[h!]
\caption{Parameter final values.} 
\label{params}
\centering
\begin{tabular}{l c c c} 
\hline
\hline
\textbf{Parameter} & \textbf{Symbol} & \textbf{Value} & \textbf{Units}\\
\hline
\hline
sampling rate & \textit{sr} & 44100 & Hz\\
window size & \textit{w} & 46 & ms\\
overlap & - & 50 & \%\\
hop length & \textit{h} & 23 & ms\\
lag & \textit{L} & 14 & s\\
pooling factor \texttt{6pool} & \textit{p} & 6 & -\\
\multirow{2}{7em}{\texttt{2pool3}} 
& $p_1$ & 2 & -\\ 
& $p_2$ & 3 & -\\
stacking parameter & \textit{m} & 2 & -\\
quantile & $\kappa$ & 0.1 & -\\
final padding & $\gamma$ & 50 & frames\\
\hline
\hline
\end{tabular}
\end{table}

\section{Dataset} 
\label{sec:dataset}

The algorithm was trained, validated and tested on a subset of the Structural Analysis of Large Amounts of Music Information (SALAMI) dataset \cite{salami}. SALAMI dataset contains 1048 double annotated pieces from which we could obtain 1006 pieces since the datasest does not provide the audio files due to copyright restrictions. For the training of the model, we used the text files of labels from annotator 1 and for the songs that were not annotated by annotator 1, we use the same text file but from annotator 2.

It is important to highlight that, as described in \cite{cohen2017music}, previous works such as \cite{ullrich2014boundary}, \cite{grill2015music} and \cite{grill2015music2} use a private non-accessible dataset of 733 songs from which 633 pieces were used for training and 100 for validation. Therefore, we re-implemented the work presented in \cite{grill2015music} but we trained it in our dataset composed by only public SALAMI pieces and annotations. We split our 1006 SALAMI audio tracks into 65\%, 15\% and 20\%, resulting in 650, 150 and 206 pieces for training, validation and testing respectively.

\subsection{Labelling Process} 
\label{labelling}

As explained in \cite{ullrich2014boundary}, it is necessary to transform the labels of the SALAMI text files into Gaussian functions so that the Neural Network can be trained correctly. 
We first set the center values of the Gaussian functions by transforming the labels in seconds into time frames as showed in Eq. \ref{eq_labels} constructing the vector $y'$ which contains the center of the gaussians and has its dimension equal to the number of labels in the text file. In Eq. \ref{eq_labels}, $\mathrm{label}_i$ are the labels in seconds extracted from SALAMI text file ``functions'' and $p1, p2, h, sr$ and $\gamma$ are defined in Table \ref{params}.

\begin{equation} 
\label{eq_labels}
y_i^{\prime} = \frac{\mathrm{label}_i}{p_1 \cdot p_2} + \frac{h \cdot sr}{\gamma}
\end{equation}

Then, we apply a gaussian function with standard deviation $\sigma$ = 0.1 and $\mu_i$ equal to each label value in Eq.\ref{eq_labels}. In Eq.\ref{gauss} we show the expression of the gaussians of the labels.

\begin{equation} 
\label{gauss}
\mathrm{gaussian\_labels_i} =  \mathbf{g}(y_i^{\prime}, \mu_i, \sigma)
\end{equation}
with
\begin{equation}
    \mathbf{g}(x,\mu,\sigma)=\frac{1}{\sqrt{2\pi}\sigma}e^{-\frac{1}{2}\left(\frac{x-\mu}{\sigma}\right)^2}
\end{equation}
where $\mu_i$ is a vector of $\frac{y_i \cdot \gamma + \frac{w}{2}}{sr}$ frames  from $i = {1}...\left[\frac{N}{p_1 \cdot p_2}\right]$.\\

To train the model, we removed the first tag from each text file due to the proximity of the first two tags in almost every file and the uselessness of the Neural Network identifying the beginning of the file. It's also worth mentioning the fact that we have resampled all the songs in the SALAMI database at a single \textit{sampling rate} of 44100Hz as showed in Table \ref{params}.

\section{Model} 
\label{sec:model}

\begin{figure}[h!]
	\centering
	\includegraphics[width=\columnwidth]{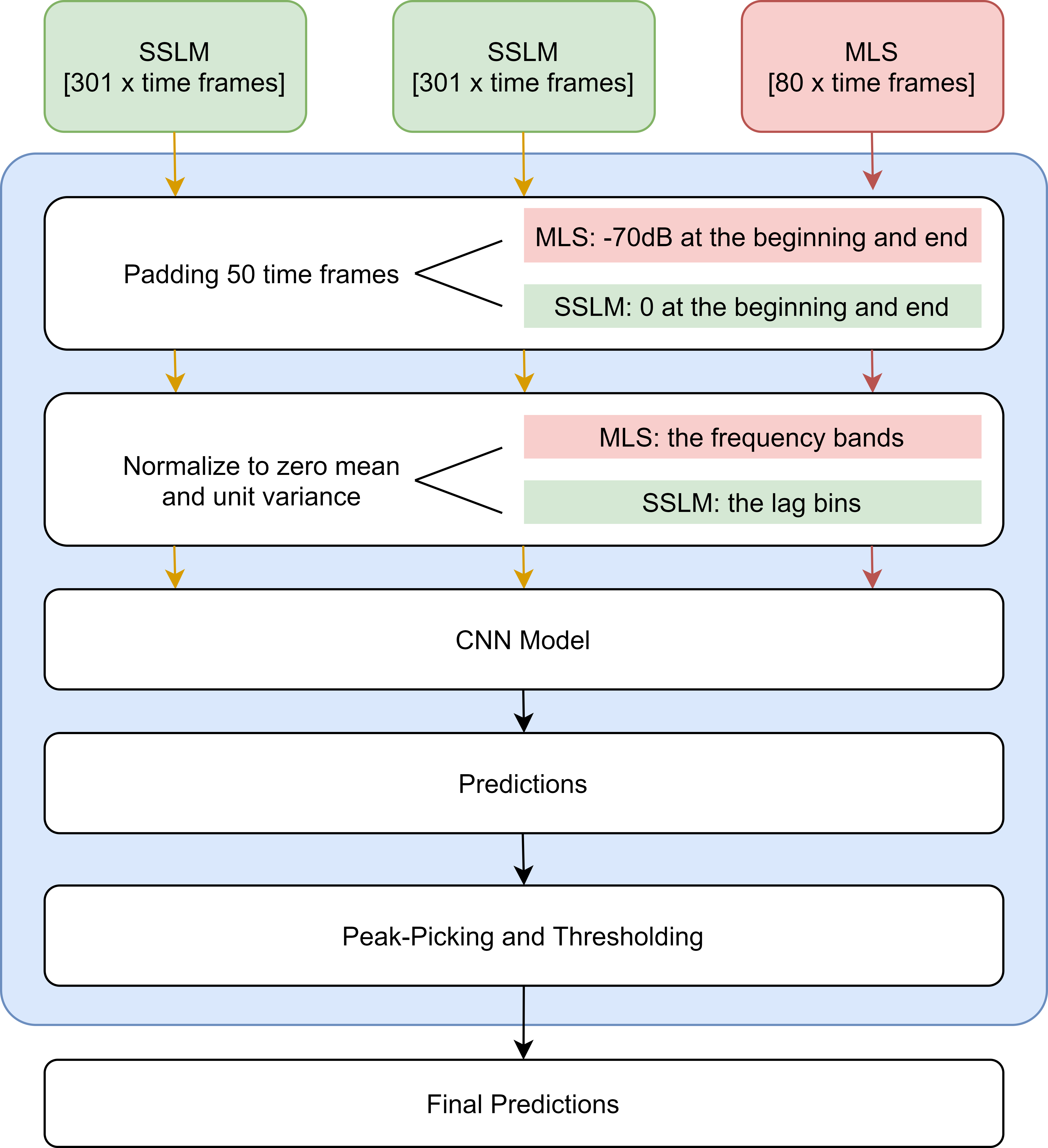}
	\caption{General block diagram of the Neural Network block in Fig. \ref{general}.}
	\label{scheme2}
\end{figure}

Our work and current methods that tackle the problem of boundary detection in MSA use neural network-based models that were originally developed for image processing tasks, in particular Convolutional Neural Networks (CNN). The model developed in this work for boundary detection is shown in Fig. \ref{scheme2}. Once the matrices of the pre-processing step are obtained, they are padded and normalized to form the input of a Convolutional Neural Network (CNN). The obtained predictions are post-processed with a peak-picking and threshold algorithm to obtain the final predictions.

\subsection{Convolutional Neural Network} 
\label{nn}

The model proposed in this paper is nearly the same than the model proposed in \cite{ullrich2014boundary} and \cite{grill2015music}, so we could compare the results and make a comparison with different input strategies as Cohen \cite{cohen2017music} did. However, we take into account more inputs combinations and with high and low dimensions in order to see the better inputs combination for the model.

\begin{figure*}[htb]
    \centering
	\includegraphics[scale=0.8]{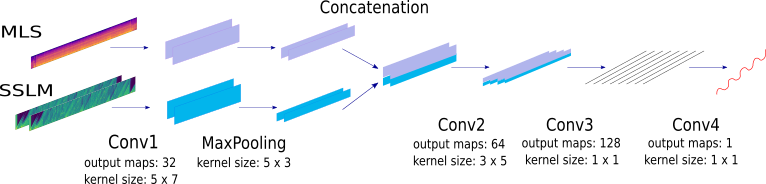}
	\caption{Schema of the Convolutional Neural Network implemented. The main parameters are presented in Table \ref{cnn_table}.}
	\label{modelCNN}
\end{figure*}

The model is composed by a CNN whose relevant parameters are shown in Table \ref{cnn_table}. The difference between this model and the model proposed in \cite{ullrich2014boundary} and \cite{grill2015music} is that our final two layers are not dense layers but convolutional layers in the time dimension because we do not crop the inputs and get a single probability value at the output, but we give the Neural Network the whole matrix and we obtain a time prediction curve at the output. The general schema of the CNN is shown in Fig. \ref{modelCNN}.

\begin{table}[htb]
\caption{CNN architecture parameters of the schema presented in Fig. \ref{modelCNN}}
    \centering
    \begin{tabular}{l l} 
    \hline
    \hline
    
    \textbf{Layer} & \textbf{Parameters}\\
    \hline
    \hline
    
    \multirow{4}{7em}{Convolution 1 + Leaky ReLU}
    & output feature maps: 32\\
    & kernel size: 5 x 7\\
    & stride: 1 x 1\\
    & padding: (5-1)/2 x (7-1)/2\\
    \hline\
    
    \multirow{3}{7em}{Max-Pooling}
    & kernel size: 5 x 3\\
    & stride: 5 x 1\\
    & padding:  1 x 1\\
    \hline
    
    \multirow{5}{7em}{Convolution 2 + Leaky ReLU}
    & output feature maps: 64\\
    & kernel size: 3 x 5\\
    & stride: 1 x 1\\
    & padding:  (3-1)/2 x (5-1)*3/2\\
    & dilation: 1 x 3\\
    \hline
    
    \multirow{4}{7em}{Convolution 3 + Leaky ReLU}
    & output feature maps: 128\\
    & kernel size: 1 x 1\\
    & stride: 1 x 1\\
    & padding:  0 x 0\\ 
    \hline\
    
    \multirow{4}{7em}{Convolution 4 + Sigmoid}
    & output feature maps: 1\\
    & kernel size: 1 x 1\\
    & stride: 1 x 1\\
    & padding:  0 x 0\\ 
    \hline
    \hline
    \end{tabular}
\label{cnn_table}
\end{table}

The parameters of the CNN model have been chosen according to previous literature \cite{grill2015music} for a fair comparison in the study of how different input features affect the performance of the MSA task. The changes that have been done from the state-of-the-art model rely on adding the dilation parameter that we use in the layers of our model, and we also changed the last layer of our implementation in comparison with previous literature models. This is because previous studies passed a segment of the SSLM trough the CNN while we pass the entire SSLM to it. 
The last layer of our implementation outputs one feature map that is passed trough a Sigmoid function which outputs the boundary probability of each time frame of the entire music piece, so the output of the model is a vector of length equal to the time frames of the input. This differs from the literature models where the output is the boundary probability of the segmented part of the input.

\subsection{Training Parameters} 
\label{cnn_params}

We trained our CNN with \textit{Binary Cross Entropy} or BCEwithLogitsLoss in Pytorch \cite{torch} as the loss function which in Pytorch implementation includes a Sigmoid activation function in the last layer of the Neural Network, a \textit{learning rate} of 0.001 and Adam optimizer \cite{kingma2015adam}. We perform \textit{early-stopping} during training to determine the best-performing model.
The SSLMs and MLS have to be passed to the GPU one by one because they have different lengths, which means that 1 song is passed forward and backward through the NN at once. However, to get more robust gradients and a more stable optimization process, the optimizer is executed with the average gradients of batchs of 10 songs. We could say that we use a batch size of 1 in terms of GPUs calls but a batch size of 10 in terms of the training.
The models were trained on a GTX 980 Ti Nvidia GPU and we used TensorboardX \cite{tensorflow2015} to graph the loss and F-score of training and validation.

\subsection{Peak-Picking} \label{peak_picking}
\begin{figure}[h]
	\centering
	\includegraphics[scale=0.65]{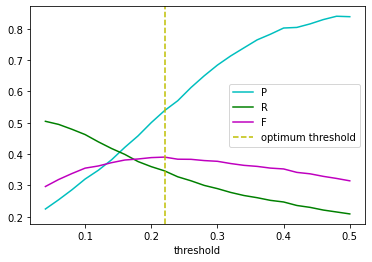}
	\caption{Threshold calculation through MLS test after 180 epochs of training with MLS.}
	\label{threshold}
\end{figure}

Peak-picking consists on selecting the peaks of the output signal of the CNN that will be identified as boundaries of the different parts of the song.
Each boundary on the output signal is considered true when no other boundary is detected within 6 seconds.
The application of a threshold helps us to discriminate boundary values that are not higher than an optimum threshold. We calculate the optimum threshold for our experiments by computing the average $F_1$ in our validation set for all possible threshold values in the range [0, 1] and then we select the highest value. Therefore, the optimum threshold is the value between [0, 1] for which the average $F_1$ is higher in our validation set. 
It is reasonable to realise that the optimum threshold value may vary when training our model with the different combination of inputs that we show in Table \ref{combination}. When we train our model with isolated inputs (see Table \ref{05s}) we compute the threshold with the MLS but we do not vary it when testing SSLMs trainings. We vary the threshold value when we train our model with different inputs combinations in order to optimize the each case of study and give the best-performing method (see Table \ref{combination}). In Fig. \ref{threshold}, we set a thresold of 0.205 for the models using only the MLS as input and for the rest of the models we used the values indicated in Table \ref{combination}. From the optimum threshold calculation, we can observe that almost all optimum threshold values for each input variant belong to $[2.05 , 2.6]$
Fig. \ref{threshold} shows Recall, Precision and F-score values (see Section \ref{evalMetrics}) of the testing dataset evaluated for each possible threshold value.

\section{Experiments and Results} \label{sec:results}

\subsection{Evaluation Metrics}
\label{evalMetrics}
MIREX's campaings use two evaluation measures which are \textit{Median Deviation} and \textit{Hit Rate}. The \textit{Hit Rate} (aslo called F-score or F-measure) is denoted by $F_\beta$, where $\beta$ = 1 is the measure most frequently used in previous works. Nieto et al. \cite{nieto2014perceptual} set a value of $\beta$ = 0.58, but the truth is that $F_1$ continues being the most used metric in MIREX works. We will later give our results for both $\beta$ values.
The \textit{Hit Rate} score $F_1$ is normally evaluated for $\pm 0.5$s and $\pm 3$s time-window tolerances, but in recent works most of the results are given only for $\pm 0.5$s tolerance which is the most restrictive one.
We test our model with MIREX algoritm \cite{raffel2014mir_eval} which give us the Precision, Recall and F-measure parameters.

\begin{equation} \label{7}
\mathrm{Precision: P = \frac{TP}{TP + FP}}
\end{equation}

\begin{equation} \label{8}
\mathrm{Recall: R = \frac{TP}{TP + FN}}
\end{equation}

\begin{equation} \label{9}
\mathrm{F}~\mathrm{measure: F_\beta = (1 + \beta^2)\frac{P \cdot R}{\beta^2 \cdot P + R}}
\end{equation}

Where: 
\begin{itemize}
  \item TP: True Positives. Estimated events of a given class that start and end at the same temporal positions as reference events of the same class, taking into account a tolerance time-window.
  \item FP: False Positives. Estimated events of a given class that start and end at temporal positions where no reference events of the same class does, taking into account a tolerance time-window.
  \item FN: False Negatives. Reference events of a given class that start and end at temporal positions where no estimated events of the same class does, taking into account a tolerance time-window.
\end{itemize}

\subsection{Results}

{\def\arraystretch{1.5}\tabcolsep=4pt
\begin{table}[h]
\caption{Results of boundaries estimation according to different pooling strategies, distances and audio features for $\pm 0.5$s and a threshold of 0.205. 
\label{05s}}
\centering
\begin{tabular}{l l c c c c}
\hline
\hline
\multicolumn{6}{c}{\textbf{Tolerance: $\pm 0.5$s and Threshold: 0.205}} \\
\hline
\hline
& \textbf{Input} & \textbf{Epochs} & \textbf{P} & \textbf{R} & $\mathbf{F_{1}}$\\
\hline
\multirow{5}{4em}{\texttt{6pool}} 
        & MLS & 180 & 0.501 & 0.359 & 0.389\\
        & $\mathrm{SSLM_{euclidean}^{MFCCs}}$ & 180 & 0.472 & 0.318 & 0.361\\
        & $\mathrm{SSLM_{cosine}^{MFCCs}}$ & 180 & 0.477
        & 0.311 & 0.355\\
        & $\mathrm{SSLM_{euclidean}^{chromas}}$ & 180 & 0.560 & 0.228 & 0.297\\
        & $\mathrm{SSLM_{cosine}^{chromas}}$ & 180 & 0.508 & 0.254 & 0.312\\
        \hline\
\multirow{2}{4em}{\texttt{2pool3}} 
        & $\mathrm{SSLM_{euclidean}^{MFCCs}}$ & 120 & 0.422 & 0.369 & 0.375\\
        & $\mathrm{SSLM_{cosine}^{MFCCs}}$ & 120 & 0.418 & 0.354 & 0.366\\
        \hline
        \hline
\multicolumn{6}{c}{\textbf{Previous works}} \\      
        \hline
\multirow{2}{4em}{\texttt{2pool3}} 
        & MLS \cite{ullrich2014boundary} & - & 0.555 & 0.458 & 0.465\\
        & $\mathrm{SSLM_{cosine}^{MFCCs}}$ \cite{grill2015music} & - & - & - & 0.430\\
        \hline
        \hline
\end{tabular}
\end{table}
}

\begin{table*}[h!]
\renewcommand{\arraystretch}{1.5}
\caption{Results of boundary estimation with tolerance $\pm 0.5$s and optimum threshold in terms of F-score, Precision and Recall. Note that results form previous works did not use the same threshold value.} \label{combination}
\centering
\begin{tabular}{p{7cm} c c c c c c c}
\hline
\hline
\multicolumn{8}{c}{\textbf{Tolerance: $\pm 0.5$s with \texttt{2pool3} matrices}}\\
\hline
\hline
\textbf{Input} & \textbf{Train Database} & \textbf{Epochs} & \textbf{Thresh.} & \textbf{P} & \textbf{R} & $\mathbf{F_{1}}$ \textbf{(std)} & $\mathbf{F_{0.58}}$\\
\hline
$\mathrm{MLS + SSLM_{euclidean}^{MFCCs}}$ & SALAMI & 140 & 0.24 & 0.441 & 0.415 & 0.402 (0.163) & 0.414\\
$\mathrm{MLS + SSLM_{cosine}^{MFCCs}}$ & SALAMI & 140 & 0.24 & 0.428 & 0.407 & 0.396 (0.158) & 0.404\\
\hline
$\mathrm{MLS + (SSLM_{euclidean}^{MFCCs} + SSLM_{euclidean}^{chromas})}$ & SALAMI & 100 & 0.24 & 0.465 & 0.400 & 0.407 (0.160) & 0.419\\
$\mathrm{MLS + (SSLM_{cosine}^{MFCCs} + SSLM_{cosine}^{chromas})}$ & SALAMI & 100 & 0.24 & 0.444 & 0.416 & 0.404 (0.166) & 0.417\\
\hline
$\mathrm{MLS + (SSLM_{euclidean}^{MFCCs}  + SSLM_{cosine}^{MFCCs})}$ & SALAMI & 100 & 0.24 & 0.445 & 0.421 & 0.409 (0.173) & 0.416\\
$\mathrm{MLS + (SSLM_{euclidean}^{chromas}  + SSLM_{cosine}^{chromas})}$ & SALAMI & 100 & 0.24 & 0.457 & 0.396 & 0.400 (0.157) & 0.420\\
\hline
$\mathrm{MLS + (SSLM_{euclidean}^{chromas}  + SSLM_{cosine}^{chromas}} +$ $\mathrm{+ SSLM_{euclidean}^{MFCCs} + SSLM_{cosine}^{MFCCs})}$ & SALAMI & 100 & 0.26 & 0.526 & 0.374 & \textbf{0.411} (0.169) & \textbf{0.451}\\
\hline
\hline
\multicolumn{8}{c}{\textbf{End-to-end previous works}} \\
\hline
\hline
$\mathrm{MLS + SSLM_{cosine}^{MFCCs}}$ \cite{grill2015music} (2015) & Private & - & & 0.646 & 0.484 & 0.523 & 0.596\\
$\mathrm{MLS + SSLM_{cosine}^{MFCCs}}$ \cite{cohen2017music} (2017) & SALAMI & - & & 0.279 & 0.300 & 0.273 (0.132) & -\\
$\mathrm{MLS + (SSLM_{cosine}^{MFCCs} + SSLM_{cosine}^{chromas})}$ \cite{cohen2017music} (2017) & SALAMI & - & & 0.470 & 0.225 & 0.291 (0.120) & -\\

\hline
\hline
\end{tabular}
\end{table*}

\begin{figure*}[h!] 
 
	\centering
	\begin{subfigure}[b]{1\textwidth}
	\centering
	    \includegraphics[scale=1.0]{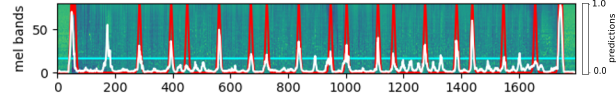}
	    \caption{CNN predictions on MLS}
	\end{subfigure}
	
	\begin{subfigure}[b]{1\textwidth}
	\centering
	    \includegraphics[scale=1.0]{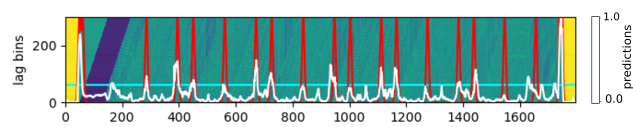}
	    \caption{CNN predictions on SSLM calculated with MFCCs and Euclidean distance with 2pool3 (best-performance SSLM input in terms of F-measure). In this case F\textsubscript{1} = 0.486 for a $\pm$0.5s tolerance.}
	\end{subfigure}
	
	\begin{subfigure}[b]{1\textwidth}
	\centering
	    \includegraphics[scale=1.0]{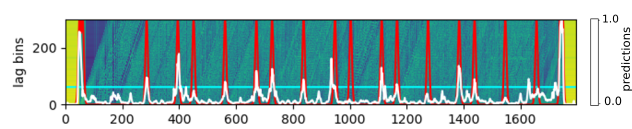}
	    \caption{CNN predictions on SSLM calculated with MFCCs and cosine distance with 2pool3. In this case F\textsubscript{1} = 0.686 for a $\pm$0.5s tolerance.}
	\end{subfigure}
	
	\begin{subfigure}[b]{1\textwidth}
	\centering
	    \includegraphics[scale=1.0]{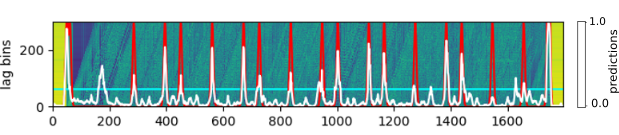}
	    \caption{CNN predictions on SSLM from MFCCs with cosine distance for model $\mathrm{MLS + (SSLM_{euclidean}^{MFCCs}  + SSLM_{cosine}^{MFCCs})}$. In this case F\textsubscript{1} = 0.75 for a $\pm$0.5s tolerance.}
	\end{subfigure}

	\caption{Boundaries predictions using CNN on different inputs obtained from the "Live at LaBoca on 2007-09-28" of DayDrug corresponding to the 1358 song of SALAMI 2.0 database. The ground truth from SALAMI annotations are the gaussians in red, the model predictions is the white curve and the threshold is the horizontal yellow line. Note that the prediction have been rescaled in order to plot them on the MLS and SSLMs images. All these images have been padded according to what is explained in the previous paragraphs and then normalized to zero mean and unit variance.} 
	\label{all_results}
\end{figure*} 

\subsubsection{Isolated Inputs: Distances, Audio Features and Pooling Strategies}

We first trained the Neural Network with each input matrix (see Fig. \ref{scheme2}) separately in order to know what input performs better. We trained the model using the MLS and SSLMs obtained from MFCCs and Chromas and applying Euclidean and cosine distances, and we also give the results for both of the pooling strategies mentioned before, \texttt{6pool} (lower resolution) and \texttt{2pool3} (higher resolution).
As mentioned in section \ref{sec:dataset}, we removed the first label of the SALAMI text files corresponding to 0.0s label. Results in terms of F score, Precision and Recall are showed in Table \ref{05s}. Note that the results showed from previous works used a different threshold value.

The best-performing input when training our model with isolated inputs is the MLS which has a $F_1$ value of 0.389 (see Table \ref{05s}). Taking only into account the \texttt{6pool} pooling strategy, regarding the SSLMs computed from audio features (MFCCs and chromas) we found that the best-performing SSLMs are the ones that are computed from the MFCCs with more than a 5\% difference with the SSLMs computed from chromas. 

According to the distance measures with which we compute the SSLMs, we found that there is not a high impact on the results when computing the SSLMs with Euclidean or cosine distances. The $F_1$ difference between the SSLMs computed with Euclidean or cosine distances is not higher than 1\%. Overall, the best-performing SSLM for the \texttt{6pool} pooling strategy is the $\mathrm{SSLM_{euclidean}^{MFCCs}}$ with a $F_1$ value of 0.361, which is a 2.8\% less than the MLS $F_!$ value of 0.389.

In view of the results in Table \ref{05s}, we can affirm that doing a max-pooling of 2, then computing the SSLMs and doing another max-pooling of 3 afterwards (\texttt{2pool3}) slightly improves the results but it does not make a high impact in the performance. The best-performing (\texttt{2pool3}) SSLM, the $\mathrm{SSLM_{euclidean}^{MFCCs}}$ has a $F_1$ value of 0.375, which is less than a 2\% of the $F_{1}$ value of 0.361 for the same SSLM but computed with the \texttt{6pool} pooling strategy. This procedure not only takes much more time to compute the SSLMs but also the training takes also much more time and it does not perform better results in terms of F-score.

In Fig. \ref{all_results} we show an example of the boundaries detection results for some of our input variants on the MLS and SSLMs. We obtained lower results than \cite{grill2015music} but higher results than \cite{cohen2017music} who tried to re-implement \cite{grill2015music}. The reasons for this difference could be that the database used by Grill and Schl\"uter \cite{grill2015music} to train their model had 733 non-public pieces. Cohen and Peeters \cite{cohen2017music}, as in our work, trained their model only with pieces from the SALAMI database, so that our results can be compared with theirs, since we trained, validated and tested our Neuronal Network with the same database (although they had 732 SALAMI pieces and we had 1006).

\subsubsection{Inputs Combination} \label{Inputs Combination}

With the higher results in Table \ref{05s} we make a combination of them as in \cite{grill2015music} and later in \cite{cohen2017music}. A summary of our results can be found in Table \ref{combination}.

The inputs combination that performs the best in \cite{cohen2017music} was $\mathrm{MLS + (SSLM_{cosine}^{MFCCs} + SSLM_{cosine}^{chromas})}$ for which $F_1$ = 0.291. We overcome that result for the same combination of inputs obtaining they obtained  a F score $F_1$ = 0.404. In spite that, previous works \cite{grill2015music} says that cosine distance performs better, we proof that in our model the Euclidean distance gives us better results. We also found that the best-performing inputs combination is $\mathrm{MLS + (SSLM_{euclidean}^{chromas}  + SSLM_{cosine}^{chromas}} + \mathrm{+ SSLM_{euclidean}^{MFCCs} + SSLM_{cosine}^{MFCCs})}$ for which $F_1$ = 0.411. There is not a huge improvement in the F-measure obtained with this combination in comparison with the results obtained with the combination of the MLS with two SSLMs, but it is still our best result.

\section{Discussion} \label{sec:discussion}

We can affirm that the best-performing input, when training the model with isolated inputs, is the Mel Spectrogram, which has a $F_1$ equal to 0.389, more than a 2\% higher than the next best-performing input respresentation, the $\mathrm{SSLM_{euclidean}^{MFCCs}}$, whose $F_1$ is equal to 0.361 (Table \ref{05s}).

We have also demonstrated that by computing a max-pooling of factor 6 at the beginning of the process not only takes much less pre-processing time but also the training of the Neural Network is faster and it does not affect the results as much as it could be expected. As an example, the $\mathrm{SSLM_{euclidean}^{MFCCs}}$ obtained with the \texttt{6pool} method has an $F_1$ value of 0.361 versus the \texttt{2pool3} method for the same input which $F_1$ is equal to 0.375.

Despite the fact that we could not replicate  some previous studies of Ullrich et al.  \cite{ullrich2014boundary} and Grill et al. \cite{grill2015music} which used nearly the same model that the one which we described in our work, we outperform the results in Cohen et al. \cite{cohen2017music} work, who also tried to re-implement the model described in the previous literature. There has to be highlighted the fact that previous studies of Ullrich et al. \cite{ullrich2014boundary} and Grill et al. \cite{grill2015music} had at their disposition a private dataset of 733 pieces that they used for training the model, and in this paper the model has been trained only with the public available dataset of SALAMI 2.0.

Adding more inputs to the model does not improve the results in a significant way and it is very time consuming, specially in our last case of study where we take 4 SSLMs in combination with the Mel Spectrogram, which has a $F_1$ value of 0.411 in contrast with the $F_1$ value of the $\mathrm{MLS + SSLM_{euclidean}^{MFCCs}}$ case which is 0.402, so the difference is less than 1\%. This leads us to suggest that the use of another neural network architecture that only uses the Mel spectrogram with a SSLM could outperform the current results.

The results obtained in this work improve those presented previously with the same database. However, the accuracy in obtaining the boundaries in musical pieces is relatively low and, to some extent, difficult to use. This makes it necessary, on the one hand, to continue studying different methods that allow a correct structural analysis of music and, on the other hand, to obtain databases that are properly labeled and contain a high number of musical pieces.
In any case, the results obtained are promising and allow us to adequately set out the bases for future work. 

\section{Conclusions} \label{sec:conclusions}

In this work we have developed a comparative study to determine the most efficient way to compute the inputs to a convolutional neural network to identify boundaries in musical pieces, combining different methods of generating SSLM matrices. In order to make the comparison and analyse the optimal way to perform the boundary detection task in MSA, different audio features and different pooling strategies have been employed, as well as the combination of different inputs to the CNN.

With an adequate combination of input matrices and pooling strategies, we obtain an accuracy F1 of 0.411 that outperforms the current one obtained under the same conditions (same input data and same datasets for training and testing). In spite of the fact that the best result is given by combining four SSLMs and the MLS, the difference in the F-measure value between our best result and experiments which require less input data and whose training time is lower, is not as high as what it could be expected. We can also affirm that current methods that have been used to date to face music boundary detection do not perform well, so MSA task needs further research because it is not solved yet. 

Future work should use new Neural Network architectures that have not been used to solve MSA yet. Architectures employed in language models from Natural Language Processing such as Transformers can lead to out-perform the actual results that are presented in this work due to the memory improvement that they provide in comparison with Long-Short Term Memory Networks (LSTMs). In the case of Transformers, the self-attention mechanism can help the model to better-process the SSMs and SSLMs matrices.
Further research, as it has been mentioned before, should also take into account to perform some data augmentation on the current public available datasets in order to have more data to train deep Neural Network models. Data augmentation, if done, should be done with pitch-shifting or by adding Gaussian noise to the inputs, but they should not use rotation or scaling techniques which affect the time distances of the input representations (horizontal axes) and thus, the structure of the music pieces.

\typeout{}
\bibliographystyle{IEEEtran}
\bibliography{refs}

\end{document}